\documentclass[aps,showpacs,pre,twocolumn,floatfix]{revtex4}
\usepackage{epsfig}
\usepackage{subfigure}
\begin{document}

\title{Molecular Dynamics Simulation on a Metallic Glass Ni$_{0.2}$Zr$_{0.8}$-System: Non-Ergodicity Parameter}
\author{A.B. Mutiara}
\affiliation{Department of Informatics Engineering, Faculty of
Industrial Technology, Gunadarma University\\Jl. Margonda Raya No. 100, Depok 16424, Indonesia\\
E-mail:amutiara@staff.gunadarma.ac.id}

\begin{abstract}
At the present paper we have computed non-ergodicity paramater from
Molecular Dynamics (MD) Simulation data after the mode-coupling
theory (MCT) for a glass transition. MCT of dense liquids marks the
dynamic glass-transition through a critical temperature $T_c$ that
is reflected in the temperature-dependence of various physical
quantities. Here, molecular dynamics simulations data of a model
adapted to Ni$_{0.2}$Zr$_{0.8}$ are analyzed to deduce $T_c$ from
the temperature-dependence of corresponding quantities and to check
the consistency of the statements. Analyzed is the diffusion
coefficients. The resulting values agree well with the critical
temperature of the non-vanisihing non-ergodicity parameter
determined from the structure factors in the asymptotic solution of
the mode-coupling theory with memory-kernels in ``One-Loop''
approximation.

\end{abstract}
\maketitle

\section{Introduction}
The transition from a liquid to an amorphous solid that sometimes
occurs upon cooling remains one of the largely unresolved problems
of statistical physics~\cite{goetze04,debenedetti01}. At the
experimental level, the so-called glass transition is generally
associated with a sharp increase in the characteristic relaxation
times of the system, and a concomitant departure of laboratory
measurements from equilibrium. At the theoretical level, it has
been proposed that the transition from a liquid to a glassy state
is triggered by an underlying thermodynamic (equilibrium)
transition~\cite{mezard99}; in that view, an ``ideal'' glass
transition is believed to occur at the so-called Kauzmann
temperature, $T_K$. At $T_K$, it is proposed that only one
minimum-energy basin of attraction is accessible to the system.
One of the first arguments of this type is due to Gibbs and
diMarzio~\cite{gibbs58}, but more recent studies using replica
methods have yielded evidence in support of such a transition in
Lennard-Jones glass formers~\cite{mezard99,coluzzi00a,grigera01}.
These observations have been called into question by experimental
data and recent results of simulations of polydisperse hard-core
disks, which have failed to detect any evidence of a thermodynamic
transition up to extremely high packing fractions~\cite{santen00}.
One of the questions that arises is therefore whether the
discrepancies between the reported simulated behavior of hard-disk
and soft-sphere systems is due to fundamental differences in the
models, or whether they are a consequence of inappropriate
sampling at low temperatures and high densities.

Different, alternative theoretical considerations have attempted
to establish a connection between glass transition phenomena and
the rapid increase in relaxation times that arises in the vicinity
of a theoretical critical temperature (the so-called
``mode-coupling'' temperature, $T_{MCT}$), thereby giving rise to
a ``kinetic'' or ``dynamic'' transition~\cite{goetze92}. In recent
years, both viewpoints have received some support from molecular
simulations. Many of these simulations have been conducted in the
context of models introduced by Stillinger and Weber and by Kob
and Andersen ~\cite{kob95a}; such models have been employed in a
number of studies that have helped shape our current views about
the glass
transition~\cite{sastry98,sciortino99,donati99,coluzzi00a,coluzzi00b,yamamoto00}.

In the full MCT, the remainders of the transition and the value of
$T_c$ have to be evaluated, e.g., from the approach of the
undercooled melt towards the idealized arrested state, either by
analyzing the time and temperature dependence in the
$\beta$-regime of the structural fluctuation dynamics
\cite{gleimkob,meyer,cum99} or by evaluating the temperature
dependence of the so-called ${\bf{g}}_m$-parameter
\cite{tei96l,tei96e}. There are further posibilities to estimates
$T_c$, e.g., from the temperature dependence of the diffusion
coefficients or the relaxation time of the final $\alpha$-decay in
the melt, as these quantities for $T>T_c$ display a critical
behaviour $|T-T_c|^{\pm \gamma}$. However, only crude estimates of
$T_c$ can be obtained from these quantities, since near $T_c$ the
critical behaviour is masked by the effects of transversale
currents and thermally activated matter transport, as mentioned
above.

On the other hand, as emphasized and applied in
\cite{barrat90,mfuchs,kobnauroth}, the value of $T_c$ predicted by
the idealized MCT can be calculated once the partial structure
factors of the system and their temperature dependence are
sufficiently well known. Besides temperature and particle
concentration, the partial structure factors are the only
significant quantities which enter the equations of the so-called
non-ergodicity parameters of the system. The latter vanish
identically for temperatures above $T_c$ and their calculation
thus allows a rather precise determination of the critical
temperature predicted by the idealized theory.

At this stage it is tempting to consider how well the estimates of
$T_c$ from different approaches fit together and whether the $T_c$
estimate from the non-ergodicity parameters of the idealized MCT
compares to the values from the full MCT. Regarding this, we here
investigate a molecular dynamics (MD) simulation model adapted to
the glass-forming Ni$_{0.2}$Zr$_{0.8}$ transition metal system.
The Ni$_x$Zr$_{1-x}$-system is well studied by experiments
\cite{kuschke,altounian} and by MD-simulations
\cite{bert1,tei92,teidimat,tei97,teib99}, as it is a rather
interesting system whose components are important constituents of
a number of multi-component 'massive' metallic glasses. In the
present contribution we consider, in particular, the $x=0.2$
compositions and concentrate on the determination of $T_c$ from
evaluating and analyzing the non-ergodicity parameter, and the
diffusion coefficients.

In the literature, similar comparison of $T_c$ estimates already exist
\cite{barrat90,mfuchs,kobnauroth} for two systems.
The studies come, however, to rather different conclusions.
From MD-simulations for a soft spheres model, Barrat et.al.\cite{barrat90}
find an agreement between the different $T_c$ estimates within about $15\%$.
On the other hand, for a binary Lennard-Jones system, Nauroth and
Kob \cite{kobnauroth} get from their MD simulations a significant
deviation between the $T_c$ estimates by about a factor of 2.
Regarding this, the present investigation is aimed at clarifying
the situation for at least one of the
important metallic glass systems.
Our paper is organized as follows: In Section \ref{SIM}, we present
the model and give some details of the computations. Section \ref{THEO}
gives  a brief discussion of some aspects of the mode coupling theory
as used here. Results of our MD-simulations and their analysis are
then presented and discussed in Section \ref{RD}.

\section{SIMULATIONS \label{SIM}}
The present simulations are carried out as state-of-the-art
isothermal-isobaric ($N,T,p$) calculations. The Newtonian
equations of $N=$ 648 atoms (130 Ni and 518 Zr) are numerically
integrated by a fifth order predictor-corrector algorithm with
time step $\Delta t$ = 2.5x10$^{-15}$s in a cubic volume with
periodic boundary conditions and variable box length L. With
regard to the electron theoretical description of the interatomic
potentials in transition metal alloys by Hausleitner and Hafner
\cite{haushafner}, we model the interatomic couplings as in
\cite{tei92} by a volume dependent electron-gas term $E_{vol}(V)$
and pair potentials $\phi(r)$ adapted to the equilibrium distance,
depth, width, and zero of the Hausleitner-Hafner potentials
\cite{haushafner} for Ni$_{0.2}$Zr$_{0.8}$ \cite{thesis}. For this
model simulations were started through heating a starting
configuration up to 2000~K which leads to a homogeneous liquid
state. The system then is cooled continuously to various annealing
temperatures with cooling rate $-\partial_tT$ = 1.5x10$^{12}$~K/s.
Afterwards the obtained configurations at various annealing
temperatures (here 1500-800 K) are relaxed by carrying out
additional isothermal annealing run at the selected temperature.
Finally the time evolution of these relaxed configurations is
modelled and analyzed. More details of the simulations are given
in \cite{thesis}.

\section{THEORY \label{THEO}}

In this section we provide some basic formulae that permit
calculation of $T_c$ and the non-ergodicity parameters $f_{ij}(q)$
for our system. A more detailed presentation may be found in
Refs.~\cite{barrat90,gotze85,bosse87,mfuchs,kobnauroth}. The
central object of the MCT are the partial intermediate scattering
functions which are defined for a binary system by \cite{bernu}
\begin{eqnarray}
F_{ij}(q,t) &=&\frac{1}{\protect\sqrt{N_{i}N_{j}}}\left\langle \rho
^{i}(q,t)\rho ^{j}(-q,0)\right\rangle  \nonumber \\
&=&\frac{1}{\protect\sqrt{N_{i}N_{j}}}\sum\limits_{\alpha
=1}^{N_{i}}\sum\limits_{\beta =i}^{N_{j}} \nonumber \\
&&\times \left\langle \exp (i{\bf{q}}\cdot
[{\bf{r}}_{\alpha }^{i}(t)-{\bf{r}}_{\beta }^{j}(0)])\right\rangle \quad,
\label{T.1}
\end{eqnarray}
where
\begin{equation}
\rho _{i }(\overrightarrow{q})=\sum\limits_{\alpha=1}^{N_{i }}e^{i
\overrightarrow{q}\cdot \overrightarrow{r}_{\alpha i }},\text{ }i =1,2
\label{T.1a}
\end{equation}
is a Fourier component  of the microscopic density of
species $i$.

The diagonal terms $\alpha=\beta$ are denoted as the incoherent intermediate
scattering function
\begin{equation}
F_{i}^{s}(q,t)=\frac{1}{N_{i}}\sum\limits_{\alpha =1}^{N_{i}}\left\langle
\exp (i{\bf{q}}\cdot [{\bf{r}}_{\alpha }^{i}(t)-{\bf{r}}_{\alpha
}^{i}(0)])\right\rangle \quad .
\label{T.2}
\end{equation}

The normalized partial- and incoherent intermediate scattering
functions are given by
\begin{eqnarray}
\Phi_{ij}(q,t)&=& F_{ij}(q,t)/S_{ij}(q) \quad ,\\
\Phi^s_{i}(q,t)&=& F_{i}^{s}(q,t) \quad ,
\end{eqnarray}
where the $S_{ij}(q)= F_{ij}(q,t=0)$ are the partial static structure factors.

The basic equations of the MCT are the set of nonlinear
matrix integrodifferential equations given by
\begin{equation}
\ddot{{\bf F}}(q,t)+{\mbox{\boldmath $\Omega $}}^2(q){\bf F}(q,t)+
\int_0^td\tau {\bf M}(q,t-\tau) \dot{{\bf F}}(q,\tau) = 0 \quad ,
\label{T.5}
\end{equation}
where ${\bf F}$ is the $2\times 2$ matrix consisting of the
partial intermediate scattering functions  $F_{ij}(q,t)$, and
the frequency matrix ${\mbox{\boldmath $\Omega $}}^2$ is given by
\begin{equation}
\left[{\mbox{\boldmath $\Omega $}}^2(q)\right]_{ij}=q^2k_B T
(x_i/m_i)\sum_{k}\delta_{ik} \left[{\bf S}^{-1}(q)\right]_{kj}\quad.
\label{T.6}
\end{equation}
${\bf S}(q)$ denotes the $2\times 2$ matrix of the
partial structure factors $S_{ij}(q)$, $x_i=N_i/N$ and $m_i$ means the
atomic mass of the species $i$.

The MCT for the idealized glass transition predicts
\cite{goetze92} that the memory kern ${\bf M}$ can be expressed at
long times by
\begin{eqnarray}
M_{ij}({\bf q},t)&=&\frac{k_B T}{2\rho m_i x_j}\int\frac{d
{\bf k}}{(2\pi)^3}
\sum_{kl}\sum_{k'l'} \nonumber \\
&& \times V_{ikl}({\bf q},{\bf k}) V_{jk'l'}({\bf q},{\bf q-k}) \nonumber \\
&& \times F_{kk'}({\bf k},t)
F_{ll'}({\bf q-k},t)\quad ,
\label{T.7}
\end{eqnarray}
where $\rho=N/V$ is the particle density and the vertex
$V_{i\alpha\beta}({\bf q},{\bf k})$ is given by
\begin{equation}
V_{ikl}({\bf q},{\bf k})=\frac{{\bf q}\cdot {\bf
k}}{q}\delta_{il} c_{ik}({\bf k})+
\frac{{\bf q}\cdot ({\bf q}-{\bf k})}{q} \delta_{ik} c_{il}
({\bf q}-{\bf k})
\label{T.8}
\end{equation}
and the matrix of the direct correlation function is defined by
\begin{equation}
c_{ij}({\bf q})=\frac{\delta_{ij}}{x_i}-
\left[{\bf S}^{-1}({\bf q})\right]_{ij} \quad .
\label{T.9}
\end{equation}

The equation of motion for $F^s_i(q,t)$ has a similar form as
eq.(\ref{T.5}), but the memory function for the incoherent
intermediate scattering
function is given by:
\begin{eqnarray}
M_{i}^{s}({\bf q},t) & = & \int \frac{d{\bf k}}{(2\pi)^3} \frac{1}{\rho}
\left(\frac{{\bf q}\cdot {\bf k}}{q}\right) (cF)_i ({\bf k},t)
\nonumber \\
&& \times F_{i}^{s}({\bf q}-{\bf k},t) ,
\label{T.10}
\end{eqnarray}
\begin{eqnarray}
(cF)_i(k,t)&=&(c_{ii}(q))^2 F_{ii}(q,t)+2c_{ii}(q)c_{ij}(q)F_{ij}(q,t)
\nonumber \\
&& +(c_{ij}(q))^2F_{jj}(q,t)\quad j\neq i \quad .
\label{T.11}
\end{eqnarray}

In order to characterize the long time behaviour of the
intermediate scattering function, the non-ergodicity parameters
${\bf f}({\bf q})$ are introduced as
\begin{equation}
f_{ij}({\bf q})=lim_{t\to \infty}\Phi_{ij}({\bf q},t) \quad .
\label{T.12}
\end{equation}
These parameters are the solution of
eqs.~(\ref{T.5})-(\ref{T.9}) at long times. The meaning of these parameters
is the following:
if $f_{ij}({\bf q})=0$, then the system is
in a liquid state with
density fluctuation correlations decaying at long times. If
$f_{ij}({\bf q})>0$, the system is in an  arrested, nonergodic state, where
density fluctuation correlations are stable for all times.
In order to compute $f_{ij}({\bf q})$, one can use the following iterative
procedure~\cite{kobnauroth}:

\begin{equation}
{\bf f}^{(l+1)}(q) = \frac{ {\bf A}(q) + {\bf B}(q)}{ {\bf C}(q) +
{\bf D}(q)} \quad, \label{T.13}
\end{equation}
where the matrix ${\bf A}(q)$, ${\bf B}(q)$,${\bf C}(q)$, ${\bf
D}(q)$, ${\bf N}(q)$ is given by
\begin{equation}
{\bf A}(q)= {\bf S}(q) \cdot {\bf N}[{\bf f}^{(l)},{\bf
f}^{(l)}](q) \cdot {\bf S} (q) \quad, \label{T.13a}
\end{equation}

\begin{equation}
{\bf B}(q)= q^{-2}|{\bf S}(q)| |{\bf N}[{\bf f}^{(l)},{\bf
f}^{(l)}](q)| {\bf S}(q)| \quad, \label{T.13b}
\end{equation}

\begin{equation}
{\bf C}(q)= q^2+Tr({\bf S}(q) \cdot {\bf N}[{\bf f}^{(l)},{\bf
f}^{(l)}](q)) \quad, \label{T.13c}
\end{equation}

\begin{equation}
{\bf D}(q)= q^{-2}| {\bf S}(q)| | {\bf N}[{\bf f}^{(l)},{\bf
f}^{(l)}](q)| \quad, \label{T.13d}
\end{equation}

\begin{equation}
N_{ij}(q)=\frac{m_i}{x_i k_B T} M_{ij}(q) \quad.
\label{T.14}
\end{equation}

This iterative procedure, indeed,  has two type of solutions,
nontrivial ones with ${\bf f}(q)>0$ and trivial solutions ${\bf f}(q)=0$.

The incoherent non-ergodicity parameter $f_i^{s}(q)$ can be
evaluated by the following iterative procedure:
\begin{equation}
q^2 \frac{f_i^{s,l+1}(q)}{1-f_i^{s,l+1}(q)} = M_i^{s}[{\bf f},
f_i^{s,l}](q)
\quad .
\label{T.15}
\end{equation}

As indicated by eq.(\ref{T.15}), computation of the incoherent
non-ergodicity parameter $f_i^s(q)$ demands that the coherent
non-ergodicity parameters are determined in advance.

\section{Results and Discussions \label{RD}}

\subsection{Partial structure factors and intermediate scattering functions}
First we show the results of our simulations concerning
the static properties of the system in terms of the partial structure factors
$S_{ij}(q)$ and partial correlation functions $g_{ij}(r)$.

To compute the partial structure factors $S_{ij}(q)$
for a binary system we use the following
definition \cite{hansen}
\begin{eqnarray}
S_{ij }(\overrightarrow{q}) &=&x_{i}\delta _{ij
}+\rho x_{i}x_{j}\int (g_{ij}(r)-1)e^{-i
\overrightarrow{q}\cdot \overrightarrow{r}}d\overrightarrow{r}
\label{E.5} \quad,
\end{eqnarray}
where
\begin{equation}
g_{ij }(\overrightarrow{r})=\frac{V}{N_{i}N_{j}}
\left\langle \sum\limits_{\alpha=1}^{N_{i}}\sum_{\beta=1,\beta\neq \alpha}^
{N_{j}}\delta ({\bf{r}}-\left| {\bf{r}}_{\alpha}(t)-
{\bf{r}}_{\beta}(t)\right| )\right\rangle  \label{E.1}
\label{E.5a}
\end{equation}
are the partial pair correlation functions.

The MD simulations yield a periodic repetition of the atomic
distributions with periodicity length $L$. Truncation of the
Fourier integral in Eq.(\ref{E.5}) leads to an oscillatory
behavior of the partial structure factors at small $q$. In order
to reduce the effects  of this truncation, we compute from
Eq.(\ref{E.5a}) the partial pair correlation functions for
distance $r$ up to $R_c=3/2L$. For numerical evaluation of
eq.(\ref{E.5}), a Gaussian type damping term is included

\begin{figure}[htbp]
\centering
\epsfig{file=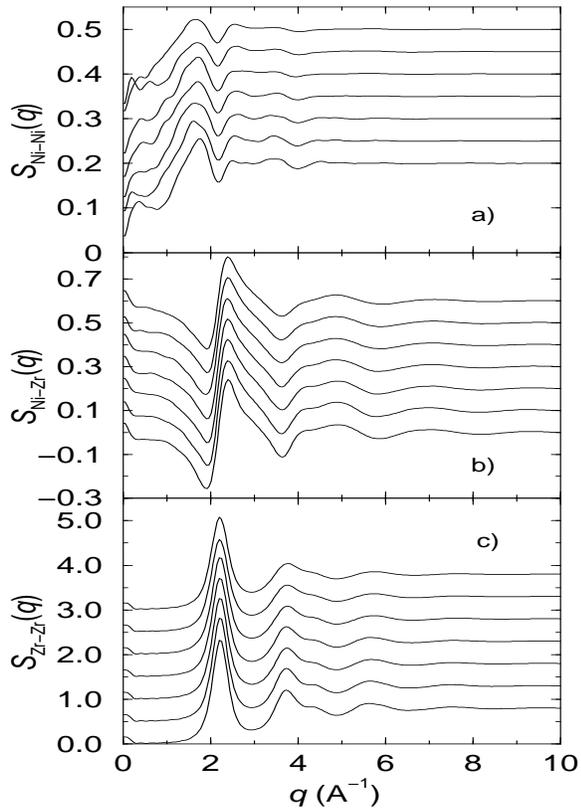,width=3in,height=4.25in}
\caption{Partial structure factors at $T=$ 1400~K, 1300~K, 1200~K,
1100~K, 1000~K, 900~K and 800~K (from top to bottom); a)
Ni-Ni-part, the curves are vertically shifted by 0.05 relative to
each other; b) Ni-Zr-part, the curves are vertically shifted by
0.1 relative to each other; and c) Zr-Zr-part, the curves are
vertically shifted by 0.5 relative to each other. } \label{fig1}
\end{figure}

\begin{figure}[htbp]
\centering \epsfig{file=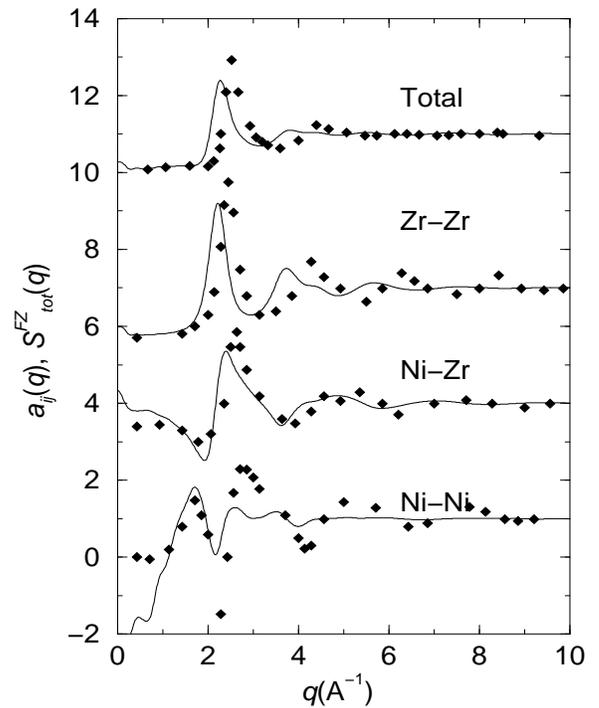,width=3in,height=3.75in}
\caption{Comparison between our MD-simulations and experimental
results \cite{kuschke} of the total Faber-Ziman structure factor
$S^{FZ}_{tot}(q)$ and the partial Faber-Ziman structur factors
$a_{ij}(q)$ for Ni$_{0.2}$Zr$_{0.8}$.} \label{fig1a}
\end{figure}

\begin{eqnarray}
S_{i j }(q)&=&x_{i }\delta _{i j }+4\pi \rho x_{i
}x_{j }\int\limits_{0}^{R_{c}}r^{2}(g_{i j }(r)-1)\frac{\sin
(qr)}{qr} \nonumber \\
&& \times \exp (-(r/R)^{2})dr
\label{E.8GS}
\end{eqnarray}
with $R=R_{c}/3$.

Fig.\ref{fig1} shows the partial structure factors $S_{ij}(q)$
versus $q$ for all temperatures investigated. The figure indicates that
the shape of $S_{ij}(q)$ depends weakly on temperature only and that, in
particular,
the positions of the first maximum and the first minimum in $S_{ij}(q)$
are more or less temperature independent.

In order to compare our calculated structure factors with experimental
ones, we have determined the Faber-Ziman partial structure factors $a_{ij}(q)$
\cite{waseda}
\begin{equation}
a_{ij}(\overrightarrow{q})=1+\rho \int (g_{ij }-1)e^{-i
\overrightarrow{q}\cdot \overrightarrow{r}}d\overrightarrow{r} \quad ,
\label{E.12}
\end{equation}
and the Faber-Ziman total structure factor $S_{tot}^{FZ}(q)$ \cite{fabzim}.
For a binary system with coherent scattering length $b_i$ of species $i$
the following relationship holds:
\begin{eqnarray}
S_{tot}^{FZ}(q) &=&\frac{1}{\left\langle b\right\rangle ^{2}}[
x_{1}^{2}b_{1}^{2}a_{11}(q)+x_{2}^{2}b_{2}^{2}a_{22}(q) \nonumber \\
&& + 2 x_{1}x_{2}b_{1}b_{2} a_{12}(q)] \quad .
\label{E.9}
\end{eqnarray}

In the evaluation of $a_{ij}(q)$, we
applied the same algorithm as for $S_{ij}(q)$. By using $a_{ij}(q)$
and with aids of the experimental data of the average scattering length
$b$ one can compute the total structure factor. Here we take $b_i$ from the
experimental data of Kuschke \cite{kuschke}. $b$ for natural Ni is 1.03
($10^{-12}$~cm) and for Zr 0.716 ($10^{-12}$~cm). Fig.\ref{fig1a} compares
the results of our simulations with the experimental
results by Kuschke \cite{kuschke} for the same alloy system at 1000 K.
There is a good agreement between the experimental and
the simulations results which demonstrates that our model is able to
reproduce the steric relations of the considered system and the
chemical order, as far is visible in the partial structure factors.

\subsection{Non-ergodicity parameters}

The non-ergodicity parameters are defined over Eq.(\ref{T.12}) as
a non-vanishing asymptotic solution of the MCT-eq.(\ref{T.5}).
Phenomenologically, they can be estimated by creating a master
curve from the intermediate scattering functions with fixed
scattering vector $q$ at different  temperatures. The master
curves are obtained by plotting the scattering functions
$\Phi(q,t)$ as function of the normalized time $t/\tau_\alpha$.

\begin{figure}[htbp]
\epsfig{file=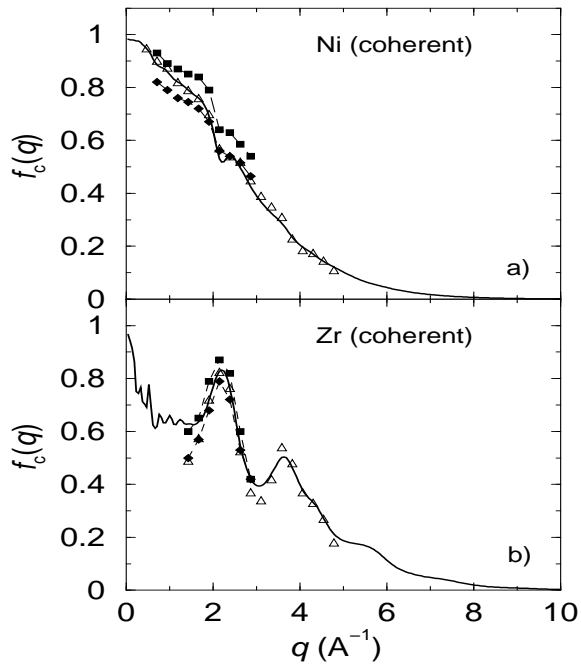,width=3in,height=3.5in}
\caption{Non-ergodicity parameter $f_{cij}$ for the coherent
intermediate scattering functions as solutions of eqs. (\ref{T.6})
and (\ref{T.7})(solid line), KWW-parameter $A(q)$ of the master
curves (diamond), Von Schweidler-parameter $f_c(q)$ of the master
curves (square), and KWW-parameter $A(q)$ for $\Phi_{ij}(q)$ at
1100 K (triangle up); a) Ni-Ni-part and b) Zr-Zr-part. }
\label{fig4a}
\end{figure}

\begin{figure}[htbp]
\epsfig{file=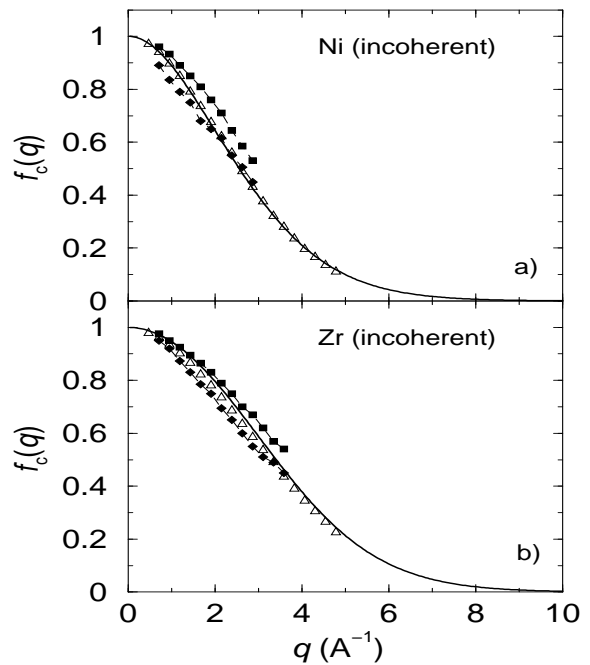,width=3in,height=3.5in} \caption{The
same  as fig.\ref{fig4a} but for the incoherent intermediate
scattering function; a) Ni-part and b) Zr-part. } \label{fig4b}
\end{figure}

Fig.~\ref{fig4a} presents the estimated $q$-dependent
non-ergodicity parameters from the coherent scattering functions
of Ni and Zr, Fig.~\ref{fig4b} those from the incoherent
scattering functions. In Fig.~\ref{fig4a} and \ref{fig4b} are also
included the deduced Kohlrausch-Williams-Watts amplitudes A(q)
from the master curves and from the intermediate scattering
functions at T=1100 K. (The further fit-parameters can be found in
\cite{thesis}.)

In order to compute the non-ergodicity parameters $f_{ij}(q)$
analytically, we followed for our binary system the
self-consistent method as formulated by Nauroth and Kob
\cite{kobnauroth} and as sketched in Section III.A. Input data for
our iterative determination of $f_{ij}(q) = F_{ij}(q,\infty)$ are
the temperature dependent partial structure factors $S_{ij}(q)$
from the previous subsection. The iteration is started by
arbitrarily setting $F_{Ni-Ni}(q,\infty)^{(0)}=0.5 S_{Ni-Ni}(q)$,
$F_{Zr-Zr}(q,\infty)^{(0)}=0.5 S_{Zr-Zr}(q)$,
$F_{Ni-Zr}(q,\infty)^{(0)}=0$.

For $T > 1200$~K we always obtain the trivial solution $f_{ij}(q)
= 0$ while at $T = 1100$~K and below we get stable non-vanishing
$f_{ij}(q)>0$. The stability of the non-vanishing solutions was
tested for more than 3000 iteration steps. From this results we
expect that $T_c$ for our system lies between 1100 and 1200 K. To
estimate $T_c$ more precisely, we interpolated $S_{ij}(q)$ from
our MD data for temperatures between 1100 and 1200 K by use of the
algorithm of Press et.al. \cite{press}. We observe that at $T =
1102$~K a non-trivial solution of $f_{ij}(q)$ can be found, but
not at $T = 1105$~K and above. It means that the critical
temperature $T_c$ for our system is around 1102 K. The non-trivial
solutions $f_{ij}(q)$ for this temperature shall be denoted the
critical non-ergodicity parameters $f_{cij}(q)$. They are included
in Fig.~\ref{fig4a}. As can be seen from Fig.~\ref{fig4a}, the
absolute values and the $q$-dependence of the calculated
$f_{cij}(q)$ agree rather well with the estimates from the
scattering functions master curve and, in particular, with the
deduced Kohlrausch-Williams-Watts amplitudes $A(q)$ at 1100 K.

By use of the critical non-ergodicity parameters $f_{cij}(q)$, the
computational procedure was run to determine the critical
non-ergodicity parameters $f^s_{ci}(q)$ for the incoherent
scattering functions at $T = 1102$~K . Fig.~\ref{fig4b} presents
our results for so calculated $f^s_{ci}(q)$. Like Fig.~\ref{fig4a}
for the coherent non-ergodicity parameters, Fig.~\ref{fig4b}
demonstrates for the $f^s_{ci}(q)$ that they agree well with the
estimates from the incoherent scattering functions master curve
and, in particular, with the deduced Kohlrausch-Williams-Watts
amplitudes $A(q)$ at 1100 K.

\subsection{Diffusion-coeffient}

From the simulated atomic motions in the computer experiments, the
diffusion coefficients of the Ni and Zr species can be determined
as the slope of the atomic mean square displacements in the
asymptotic long-time limit
\begin{equation}
D_{i}(T)=\lim\limits_{t\rightarrow \infty }\frac{(1/N_{i})\sum\limits_{%
\alpha =1}^{N_{i}}\left| \mathbf{r}_{\alpha }(t)-\mathbf{r}_{\alpha
}(0)\right| ^{2}}{6t} \quad.
\end{equation}

\begin{figure}[htbp]
\epsfig{file=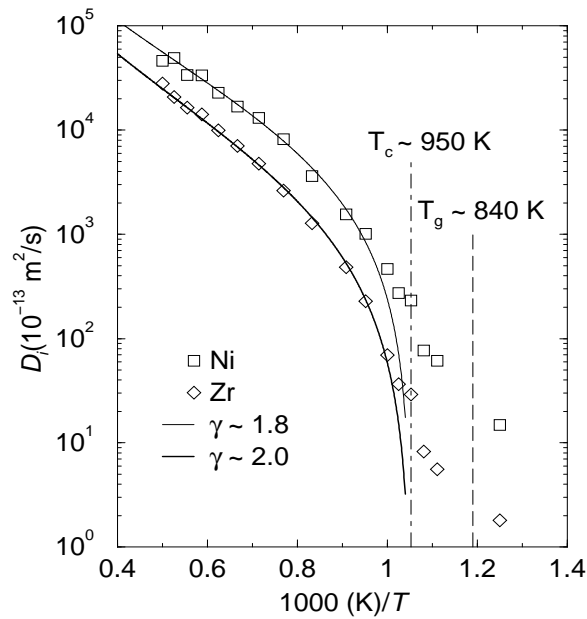,width=3in,height=3.25in}
\caption{Diffusion coefficients $D_i$ as a function of $1000/T$.
Symbols are MD results for Ni (square) and Zr (diamond); the full
line are a power-law approximation for Ni and for Zr. resp..}
\label{fig9}
\end{figure}

Fig.~\ref{fig9} shows the thus calculated diffusion coefficients of our
Ni$_{0.2}$Zr$_{0.8}$ model for the temperature range between 800 and 2000 K.
At temperatures above approximately 1250 K, the diffusion coefficients for
both species run parallel with temperature in the Arrhenius plot, indicating a fixed
ratio $D_{Ni}/D_{Zr}\approx  2.5$ in this temperature regime. At lower
temperatures, the Zr atoms have a lower mobility than the Ni atoms,
yielding around 900 K a value of about 10 for $D_{Ni}/D_{Zr}$.
That means, here the Ni atoms carry out a rather rapid motion within
a relative immobile Zr matrix.

According to the MCT, above $T_c$ the diffusion coefficients follow
a critical power law
\begin{equation}
D_{i}(T)\sim (T-T_{c})^{\gamma }, \text{ for }T > T_c \label{MA.1}
\end{equation}
with non-universal exponent $\gamma$ \cite{kob95a,hansenyip}. In
order to estimate $T_c$ from this relationship, we have adapted
the critical power law by a least mean squares fit to the
simulated diffusion data for 1050 K and above. The results of the
fit are included in Fig.~\ref{fig9} by dashed lines. According to
this fit, the system has a critical temperature of 950 K. The
parameters $\gamma$ turn out as 1.8 for the Ni subsystem and 2.0
for the Zr system.

\section{Conclusion}
The results of our MD-simulations show that our system behave so
as predicted by MCT in the sense that the diffusion coefficients
follow the critical power law. After analizing this coefficient we
found that the system has critical temperature of 950 K.

Our analysis of the ergodic region (~$T>T_c$~) and of the
non-ergodic region (~$T<T_c$~) lead to $T_c$-estimations which
agree each other within 10~$\%$. These $T_c$-estimations are also
in acceptable compliance with the $T_c$-estimation from the
dynamic phenomenons. Within the scope of the precision of our
analysis, the critical temperatur $T_c$ of our system is about
1000~K.

\end{document}